\documentclass[aps,prc,letterpaper,11pt,twoside,tightenlines,nofootinbib,showpacs,preprint]{revtex4-1}
\begin{document}

\title{$X(3872)$ and its Partners in the Heavy Quark Limit of QCD}
\author{C. Hidalgo-Duque\footnote{carloshd@ific.uv.es}}
\author{J. Nieves\footnote{jmnieves@ific.uv.es}}
\author{A. Ozpineci\footnote{ozpineci@metu.edu.tr (On sabbatical leave from Physics Department, Middle East Technical University, Ankara, Turkey)}} 

\affiliation{Instituto de Física Corpuscular (centro mixto CSIC-UV), Institutos de Investigación de Paterna, Aptdo. 22085, 46071, Valencia, Spain}
\author{V. Zamiralov\footnote{zamir@depni.sinp.msu.ru}}
\affiliation{Skobeltsyn Institute of Nuclear Physics, Lomonosov MSU, Moscow, Russia}
\begin{abstract}
In this letter, we propose interpolating currents for the $X(3872)$ resonance, and show that, in the Heavy Quark limit of QCD, the $X(3872)$ state should have degenerate partners, independent of its internal structure. Magnitudes of possible $I=0$ and $I=1$ components of the $X(3872)$ are also discussed.
\end{abstract}
\maketitle
\thispagestyle{empty}
\section{Introduction}
Although it has been almost 10 years since its observation by the BELLE collaboration \cite{PhysRevLett.91.262001}, the nature of the $X(3872)$ meson still remains a puzzle in charmonium physics. After its first observation, it has also been confirmed by \cite{PhysRevLett.93.072001, PhysRevLett.93.162002, PhysRevD.71.071103, Collaboration:2013rt}. Its $J^{PC}$ quantum numbers have not been established, but the analysis of the $X(3872) \rightarrow J/\Psi\pi\pi$ decays are consistent with the assignments $J^{PC} =1^{++}$ and $J^{PC} = 2^{-+}$ \cite{PhysRevD.84.052004,PhysRevLett.98.132002}, the former being slightly more favorable \cite{Brambilla:2010cs}.

As to its decay modes, besides the $J/\Psi\pi^+\pi^-$ final state, the decay into $J/\Psi\pi^+\pi^-\pi^0$ was also observed \cite{Abe:2005ix,Amo-Sanchez:2010fk}. The analysis of the data indicates that the decays into final states containing two and three pions decays proceed through the intermediate states $J/\Psi \rho$ and $J/\Psi\omega$, respectively. Branching ratios of the decays $X(3872) \rightarrow J/\Psi \rho$ and $X(3872) \rightarrow J/\Psi \omega$ are found to be comparable. This fact indicates that there is a large isospin violation in the decays of $X(3872)$ meson \cite{Tornqvist:2004qy}.

Its puzzling decay patterns makes it hard to fit the $X(3872)$ meson within the traditional mesons. 
Closeness of its mass to the $D^0D^{*0}$ threshold ($m(X(3872)) - [m(D^0)+m(D^{*0})] = -0.42\pm0.39$ MeV \cite{Brambilla:2010cs}) hints to the possibility that it might be a weakly bound state of a $D$ and a $\bar D^{*}$ with possible admixtures from $J/\Psi \omega$ and $J/\Psi \rho$ states \cite{Swanson:2003tb} and/or charmonium \cite{Takizawa:2012fk,Narison:2011uq,Ortega:2010qq}. In the $D\bar D^*$ model, the isospin breaking effects arise due to the mass difference between the $D^0D^{*0}$ system and its charged counterpart $D^+\bar D^{*-}$ system \cite{Tornqvist:2004qy,Gamermann:2009fv,Gamermann:2009uq}.  

Several models have been proposed to study the properties of $X(3872)$ using the heavy quark effective field theory approach (see e.g., \cite{Colangelo2007166,Hidalgo-Duque:2013fk,Nieves:2012fk}). One of the results obtained in \cite{Hidalgo-Duque:2013fk,Nieves:2012fk} using heavy quark symmetry states that if $X(3872)$ is a bound state of a $D$ and a $\bar D^*$, then it  should have  degenerate partners in the heavy quark limit. In this work, the question of partners of the $X(3872)$ meson will be addressed directly using QCD in the heavy quark limit. In section \ref{partners}, an interpolating current for the $X(3872)$ is proposed, and its correlator  is studied to prove that there should be three other degenerate partners of the $X(3872)$. Section \ref{isospin} is devoted to the study of isospin mixing in the $X(3872)$ mesons, and finally in section \ref{conclusion}, our results are summarized.
\section{$X(3872)$ Partners}
\label{partners}
The fundamental object to study the properties of hadrons in a field theory framework is a correlation function of the form
\begin{eqnarray}
\Pi = i \int d^4x e^{i p x} \langle 0 \vert {\cal T} j(x) j^\dagger(0) \vert 0 \rangle
\end{eqnarray}
where $j(x)$ is a suitably chosen interpolating current that can create the hadron of interest from the vacuum,  $p$ is the momentum of the state created by $j(x)$, and ${\cal T}$ is the time ordering operator.
This correlation function can be written in terms of hadronic properties, and is the main object of interest in many non-perturbative methods such as QCD sum rules, lattice QCD, AdS/QCD, etc.

In the molecular representation of $X(3872)$ as a bound state of $D$ and $D^*$, the $J^{PC} =1^{++}$ combination can be written as
$$
\vert X(3872) \rangle = \frac{1}{\sqrt2} \left( \vert D \bar D^* \rangle - \vert \bar D D^* \rangle \right)
$$
In terms of the quarks, one can express this state as
\begin{eqnarray}
\vert X(3872) \rangle = \vert (S_{c\bar c}=1 ;S_{q\bar  q}=1)J=1 \rangle
\label{spincoupling}
\end{eqnarray}
i.e., the $c$ and $\bar c$ quarks and $q$ and $\bar q$ quarks, each combine in a spin-1 state, and the two pairs combine to give a state with overall $J=1$.
Motivated by
the latter picture, one can choose a current of the form
\begin{eqnarray}
j_{\alpha \beta}^q = \bar Q^a \gamma_\alpha Q^b \bar q^b \gamma_\beta q^a \label{current1}
\end{eqnarray}
where $Q$ is a heavy quark ($c$ in the case of $X(3872)$), $q$  is any quark different from $Q$, $a$ and $b$ are color indices written in a way to assure colorless $D$ and $D^*$ states. Although we have chosen the two quark fields $q$ and $\bar q$ appearing in Eq. (\ref{current1}) to be identical for simplicity, the following discussion holds even if they are different quark fields. (The same is true for the heavy quark $Q$, with obvious modifications in the following discussions).

Note that one can rewrite the color factors of the $q\bar q$ and $Q \bar Q$ as
$$\bar Q^a \gamma_\alpha Q^b = \left( \bar Q^a \gamma_\alpha Q^b - \frac{\delta^{ab}}{3} \bar Q^c \gamma_\alpha Q^c \right) +  \frac{\delta^{ab}}{3} \bar Q^c \gamma_\alpha Q^c $$
and a similar expansion for the other quark-antiquark current.
Here the first term is a color octet combination and the last term is a color singlet. In this form it can be seen that
this current also has a non-trivial coupling to the $J/\Psi\omega$ and/or $J/\Psi\rho$ components of the $X(3872)$ state \cite{Tornqvist:2004qy}.

In the following discussion, we will drop the superscript $q$ until section \ref{isospin}. The current in Eq. (\ref{current1})  has even charge conjugation $C=+1$. 
(An alternative current with $C=+1$ is $j_{\alpha \beta} = \bar Q^a \sigma_{\alpha \delta} Q^b \bar q^b \sigma^\delta_{~\beta} q^a$. This current yields to the same final results.)
The product of two vector currents can be written as a sum of irreducible representations of the Lorentz group upon using the projection operators:
\begin{eqnarray}
{\cal P}^2_{\mu\nu;\bar \mu \bar \nu} &=& \frac12 \left( g_{\mu \bar \mu} g_{\nu \bar \nu} + g_{\mu \bar \nu} g_{\nu \bar \mu} - \frac12 g_{\mu \nu} g_{\bar \mu \bar \nu} \right) \label{projection2}
\\
{\cal P}^1_{\mu \nu; \bar \mu \bar \nu} &=& \frac12 \left( g_{\mu \bar \mu} g_{\nu \bar \nu} - g_{\mu \bar \nu} g_{\nu \bar \mu} \right) \label{projection1}
\\
{\cal P}^0_{\mu \nu;\bar \mu \bar \nu} &=&\frac14 g_{\mu \nu} g_{\bar \mu \bar \nu} \label{projection0}
\end{eqnarray}
as
\begin{eqnarray}
j^2_{\mu \nu} &=& {\cal P}^2_{\mu \nu; \bar \mu \bar \nu} j^{\bar \mu \bar \nu} \label{spin2current}
\\
j^1_{\mu \nu} &=& {\cal P}^1_{\mu \nu; \bar \mu \bar \nu} j^{\bar \mu \bar \nu} \label{spin1current}
\\
j^0_{\mu \nu} &=& {\cal P}^0_{\mu \nu;\bar \mu \bar \nu} j^{\bar \mu\bar \nu} \label{spin0current}
\end{eqnarray}
where the superscript denotes the largest spin that a particle that couples to the corresponding current can have.

Let us first consider the correlation function constructed from the current given in Eq. (\ref{current1}):
\begin{eqnarray}
\Pi_{\alpha \beta; \gamma \delta} &=& i \int d^4 x e^{i p x} \langle 0 \vert {\cal T} j_{\alpha \beta}(x) j^\dagger_{\gamma \delta}(0) \vert 0 \rangle
\nonumber \\
&=& i \int d^4 x e^{i p x} \langle 0 \vert {\cal T}\bar Q^a(x) \gamma_\alpha Q^b(x) \bar q^b(x) \gamma_\beta q^a(x)
\nonumber \\
&&~~~~~~~~~~~~~~~~\bar Q^d(0) \gamma_\gamma Q^e(0) \bar q^e(0) \gamma_\delta q^d(0) \vert 0 \rangle
\nonumber \\
\label{correlationfunction}
\end{eqnarray}

In the heavy quark limit, the momentum of the state created by the current is written as $p=2 m_Q v + k$, where $v$ is the common velocity of the heavy quarks and $k$ is the residual momentum of the system. The field corresponding to the quark Q can be decomposed as:
\begin{eqnarray}
Q(x) = e^{- i m_Q v x} h_Q^{(+)}(x) + e^{i m_Q v x} h_{\bar Q}^{(-)}(x)
\label{resolutions}
\end{eqnarray}
where $h_Q^{(+)}(x)$($h_{\bar Q}^{(-)}(x)$) contains the  positive (negative) frequency components, i.e. $h_Q^{(+)}(x)$ contains the annihilation operators for the quark and $h_{\bar Q}^{(-)}(x)$ contains the creation operators for the antiquark. Substituting Eq. (\ref{resolutions}) into Eq. (\ref{correlationfunction}), and taking the heavy quark limit, the correlation function becomes:
\begin{eqnarray}
\Pi_{\alpha \beta; \gamma \delta} =
i \int d^4 x e^{i k x} \langle 0 \vert {\cal T}\overline {h_{\bar Q}^{(-)a}}(x) \gamma_\alpha h_Q^{(+)b}(x)\bar q^b(x)   \gamma_\beta q^a(x)
\nonumber \\
\overline{h_Q^{(+)d}}(0) \gamma_\gamma h_{\bar Q}^{(-)e}(0) \bar q^e(0) \gamma_\delta q^d(0) \vert 0 \rangle
\nonumber \\
\label{corrheavy}
\end{eqnarray}
Note that in obtaining Eq. (\ref{corrheavy}), terms that contain an  exponential factor with infinite oscillation frequency have been neglected.
 
In the heavy quark limit, the spin of the heavy quark decouples from the theory. Hence all the gamma matrices multiplying the heavy quarks can be factored out of the correlation function:
\begin{eqnarray}
\Pi_{\alpha \beta; \gamma\delta } &=& \mbox{Tr} \left[\gamma_\alpha \frac{1+\not\!v}{2} \gamma_\gamma \frac{1-\not\!v}{2} \right]
\left( {\cal R}_1 g_{\beta \delta} + {\cal R}_2 v_\beta v_\delta \right)
\nonumber \\
&=& 2 \left(g_{\alpha \gamma} - v_\alpha v_\gamma\right)\left( {\cal R}_1 g_{\beta \delta} + {\cal R}_2 v_\beta v_\delta \right)
\label{factorgamma}
\end{eqnarray}
where the most general decomposition of the remaining part of the correlation function is written in terms of Lorentz invariant functions ${\cal R}_1$ and ${\cal R}_2$. Applying the projection operators Eqs. (\ref{projection2}-\ref{projection0}), the correlation functions for the currents Eqs. (\ref{spin2current}-\ref{spin1current}) can be obtained as (in this work we will not need the correlation function for the current given in Eq. (\ref{spin0current})):
\begin{eqnarray}
\Pi^{2}_{\mu \nu;\bar \mu \bar \nu} &=& 2 {\cal P}_{2\mu \nu; \bar \mu \bar \nu} {\cal R}_1 -
\frac18 {\cal R}_1 \left( g_{\mu \nu} - 4 v_\mu v_\nu \right) \left( g_{\bar \mu \bar \nu} - 4 v_{\bar \mu} v_{\bar \nu} \right)
\nonumber \\ &&
+\frac12 ({\cal R}_2 - {\cal R}_1) \left[
   v_\nu v_{\bar \nu} \left(g_{\mu \bar \mu} - v_\mu v_{\bar \mu} \right)
+ v_{\bar \mu} v_\nu \left(g_{\mu \bar \nu} - v_\mu v_{\bar \nu} \right)
\right. \nonumber \\ && \left.
+ v_\mu v_{\bar \nu} \left(g_{\bar \mu \nu} - v_{\bar \mu} v_\nu \right)
+v_\mu v_{\bar \mu} \left(g_{\nu \bar \nu} - v_{\nu} v_{\bar \nu} \right)
 \right] 
 \label{alternative2p} \\
 \Pi^{1}_{\mu \nu; \bar \mu \bar \nu} &=& 
 {\cal R}_1 \left( g_{\mu \bar \mu} g_{\nu \bar \nu} - g_{\mu \bar \nu} g_{\nu \bar \mu} \right) 
 \nonumber \\ &&
 -\frac12 ( {\cal R}_1 - {\cal R}_2 ) \left( v_\nu v_{\bar \nu} g_{\mu \bar\mu} - v_{\bar \mu} v_\nu g_{\mu \bar \nu} - v_\mu v_{\bar \nu} g_{\nu \bar \mu} + v_\mu v_{\bar \mu} g_{\nu \bar \nu} \right)
 \nonumber \\
 \label{alternative1p}
\end{eqnarray}

To calculate the expression for the correlation function in terms of the hadronic states, note the $j_{\mu \nu}^1$ current, being anti-symmetric under its indices,  couples only to $J^P=1^+$ and $J^P=1^-$ states, whereas the $j_{\mu \nu}^2$ current couples to states with $J^P=2^+$, $J^P=0^+$ and $J^P=1^-$ states.

The couplings of the $j_{\mu \nu}^1$ current to the $J^P=1^+$ and $J^P=1^-$ particles can be written as
\begin{eqnarray}
\langle 1^- \vert j_{\mu \nu}^{1} \vert 0 \rangle &=& A \left(v_\mu \epsilon_\nu - v_\nu \epsilon_\mu\right)
\nonumber \\
\langle 1^+ \vert j_{\mu \nu}^{1} \vert 0 \rangle &=& A' \epsilon_{\mu \nu \alpha \beta} v^\alpha \epsilon'^\beta
\end{eqnarray}
where $A$($A'$) and $\epsilon^{(')}$ are the coupling strength and the polarization vector of the (axial)vector particle. Summing over the polarizations of the (axial) vector using
\begin{eqnarray}
\sum_{polarization} \epsilon_\mu \epsilon^*_\nu = - \left(g_{\mu \nu} - v_\mu v_\nu \right)
\label{vecpolsum}
\end{eqnarray}
it is seen that the correlation function can be written as
\begin{eqnarray}
\Pi^1_{\mu\nu;\bar\mu\bar\nu} &=& - \frac{A^2}{kv - \Lambda_{1^-}} \left[ \left( v_\mu v_{\bar \mu} g_{\nu \bar \nu} - v_\mu v_{\bar \nu} g_{\nu \bar \mu} - v_\nu v_{\bar \mu} g_{\mu \bar \nu} + v_\nu v_{\bar \nu} g_{\mu \bar \mu} \right)\right]
\nonumber \\
&& + \frac{A'^2}{kv - \Lambda_{1^+}}\left[ \left( g_{\mu \bar \mu} g_{\nu \bar \nu} - g_{\mu \bar \nu} g_{\nu \bar \mu} \right) -
 \left( v_\nu v_{\bar \nu} g_{\mu \bar\mu} - v_{\bar \mu} v_\nu g_{\mu \bar \nu} - v_\mu v_{\bar \nu} g_{\nu \bar \mu} + v_\mu v_{\bar \mu} g_{\nu \bar \nu} \right) \right]
\nonumber \\
&=&   \frac{A'^2}{kv - \Lambda_{1^+}} \left( g_{\mu \bar \mu} g_{\nu \bar \nu} - g_{\mu \bar \nu} g_{\nu \bar \mu} \right) 
\nonumber \\
&& - \left( \frac{A^2}{kv - \Lambda_{1^-}}  + \frac{A'^2}{kv - \Lambda_{1^+}} \right)
\left( v_\nu v_{\bar \nu} g_{\mu \bar\mu} - v_{\bar \mu} v_\nu g_{\mu \bar \nu} - v_\mu v_{\bar \nu} g_{\nu \bar \mu} + v_\mu v_{\bar \mu} g_{\nu \bar \nu} \right)
\end{eqnarray}
where a sum over all states with $J^P=1^+$ and $J_P=1^-$ should be understood. In this equation, the constants $\Lambda_{J^P}$  are defined as $\Lambda_{J^P} = \lim_{m_Q \rightarrow 0} (m_{J^P} - 2 m_Q)$ with $m_{J^P}$ denoting the mass of the corresponding state. In the molecular picture of the these states, $\Lambda_{J^P}$ corresponds to the interaction energy of the mesons at zero spatial separation \cite{Detmold:2007wk}.

Comparing with Eq. (\ref{alternative1p}), the ${\cal R}_1$ and ${\cal R}_2$ functions can be identified as:
\begin{eqnarray}
{\cal R}_1 &=&  \frac{A'^2}{kv-\Lambda_{1^+}}
\nonumber \\
{\cal R}_1 -{\cal R}_2 &=& 2 \frac{A^2}{kv - \Lambda_{1^-}} +2 \frac{A'^2}{kv-\Lambda_{1^+}}
\label{aresults1p}
\end{eqnarray}

On the other hand, one can also find an expression for the functions ${\cal R}_1$ and ${\cal R}_2$ using Eq. (\ref{alternative2p}). The coupling of the $J^P=2^+$, $J^P=0^+$ and $J^P=1^-$ states to the $j^2_{\mu \nu}$ currents can be written as
\begin{eqnarray}
\langle 2^+ \vert j_{\mu \nu}^{2} \vert 0 \rangle &=& D \epsilon_{\mu \nu}
\nonumber \\
\langle 0^+ \vert j_{\mu\nu}^{2} \vert 0 \rangle &=& C \left( g_{\mu \nu} - 4 v_\mu v_\nu \right).
\nonumber \\
\langle 1^- \vert j_{\mu \nu}^{2} \vert 0 \rangle
&=& F \left(\epsilon_\mu v_\nu + \epsilon_\nu v_\mu- \frac12 g_{\mu \nu} \epsilon \cdot v \right)
\end{eqnarray}
where $\epsilon_\mu$ and $\epsilon_{\mu \nu}$ are the polarizations of the vector and tensor particles respectively, satisfying $v^\mu \epsilon_\mu = 0$, $\epsilon_\mu \epsilon^{\mu*} = -1$, $v^\mu \epsilon_{\mu \nu} = 0$, $\epsilon_{\mu \nu} = \epsilon_{\nu \mu}$, $\epsilon_{\mu \nu} g^{\mu \nu}=0$, $\epsilon_{\mu \nu} \epsilon^{\mu \nu*} = 1$.

Inserting a complete set of these states into the correlation function and summing over the polarizations upon using Eq. (\ref{vecpolsum}) for vector particles and
\begin{eqnarray}
\sum_{polarization} \epsilon_{\mu \nu} \epsilon^*_{\alpha \beta} &=& \frac12 \left[
 \left(g_{\mu \alpha} - v_\mu v_{\alpha} \right)
 \left(g_{\nu \beta} - v_\nu v_\beta \right) +
 \left(g_{\nu \alpha} - v_\nu v_{\alpha} \right)
 \left(g_{\mu \beta} - v_\mu v_\beta \right)
 \right. \nonumber \\
 &&\left.
 - \frac 23 \left(g_{\mu \nu} - v_\mu v_\nu \right) \left( g_{\alpha \beta} - v_\alpha v_\beta \right)
  \right]
\end{eqnarray}
for  the tensor particles, the correlation function of Eq. (\ref{alternative2p}) can be written as
\begin{eqnarray}
\Pi^2_{\mu \nu; \bar \mu \bar \nu} &=& 
  \frac{D^2}{kv - \Lambda_{2^+}} {\cal P}^2_{\mu\nu;\bar\mu\bar\nu} 
  \nonumber \\ &&
+ \left(\frac{C^2}{kv - \Lambda_{0^+}} - \frac{D^2}{12(kv-\Lambda_{2^+})}  \right)
 \left( g_{\mu \nu} - 4 v_\mu v_\nu \right)
\left( g_{\bar \mu \bar \nu} - 4 v_{\bar \mu} v_{\bar \nu} \right)
\nonumber \\&&
-\left( \frac{D^2/2}{kv - \Lambda_{2^+}} + \frac{F^2}{kv-\Lambda_{1^-}}   \right)
\left[
v_\nu v_{\bar \nu} \left(g_{\mu \bar \mu} - v_\mu v_{\bar \mu} \right)
+v_\nu v_{\bar \mu} \left(g_{\mu \bar \nu} - v_\mu v_{\bar \nu} \right)
\right.
\nonumber \\
&&+ \left. v_\mu v_{\bar \nu} \left(g_{\nu \bar \mu} - v_\nu v_{\bar \mu} \right)
+v_\mu v_{\bar \mu} \left(g_{\nu \bar \nu} - v_\nu v_{\bar \nu} \right)
 \right]
\end{eqnarray}
Comparing with Eq. (\ref{alternative2p}), it is seen that in this case
\begin{eqnarray}
{\cal R}_1 &=&  \frac{D^2/2}{kv - \Lambda_{2^+}}
\nonumber \\
{\cal R}_1 &=& -8 \frac{C^2}{kv - \Lambda_{0^+}} + \frac23 \frac{D^2}{kv - \Lambda_{2^+}}
\nonumber \\
{\cal R}_1-{\cal R}_2 &=&  \frac{D^2}{kv - \Lambda_{2^+}} + \frac{2F^2}{kv-\Lambda_{1^-}}
\label{aresults2p}
\end{eqnarray}
Note that, requiring the consistency of the two expression of ${\cal R}_1$ in Eq. (\ref{aresults2p}), one arrives at the equalities $\Lambda_{0^+}=\Lambda_{2^+}$ and $D^2=48 C^2$. Furthermore,  Eqs. (\ref{aresults1p}) and (\ref{aresults2p}) are consistent only if $A^2=F^2$, $2A'^2=D^2$, and $\Lambda_{2^+}=\Lambda_{1^+}$. Hence, there should be two more states, with $J^{PC}=2^{++}$ and $0^{++}$, degenerate with the $X(3872)$ resonance. 

We would like to stress here that the degeneracy obtained in this section does not make any assumption about the nature of the two quark fields $q$ appearing in the current given in Eq. (\ref{current}). Hence, these degeneracies also hold for states with $I=0$ (with or without hidden flavor), $I=1$ and  $I=1/2$.

The obtained degeneracies will be broken by finite mass effects.
The authors of \cite{Hidalgo-Duque:2013fk,Nieves:2012fk} derive an effective field theory scheme to describe $D^{(*)} \bar D^{(*)}$ molecules implementing leading order heavy quark spin symmetry constraints on the dynamics. In these works it is shown that the dynamics of the $2^{++}$ and $1^{++}$ channels are identical. However the predicted $2^{++}$ and $1^{++}$ ($X(3872)$) states are not degenerate because of the $D^*-D$ mass difference, which is a consequence of the finite value of the charm quark mass.

On the other hand, in the scheme of \cite{Hidalgo-Duque:2013fk,Nieves:2012fk}, there appears also a $0^{++}$ state degenerate with the $1^{++}$ and $2^{++}$ states mentioned above in the infinite quark mass limit. This new state is a result of the $D \bar D$ and $D^* \bar D^*$ coupled channel dynamics (see Eqs. (18-21) of \cite{Nieves:2012fk}). This state is similar to the $1^{++}$ and $2^{++}$ states in the sense that in this state the heavy quarks are coupled to spin-1 and the light quark are coupled to spin-1  as shown in Eq. (\ref{spincoupling}). However, for finite charm quark masses, coupled channel effects are subleading in the expansion proposed in \cite{Hidalgo-Duque:2013fk,Nieves:2012fk}. When these
are neglected, the dynamics of the $0^{++}$ state predicted in \cite{Hidalgo-Duque:2013fk,Nieves:2012fk} is different to that governing the $1^{++}$ and $2^{++}$ sectors \cite{Hidalgo-Duque:2013fk,Nieves:2012fk}.

Upto now, the discussion has been limited to the $C=+1$ states. In \cite{Hidalgo-Duque:2013fk,Nieves:2012fk}, it was also observed that there is another state with the same binding energy as the $0^{++}$, $1^{++}$ and $2^{++}$ states and with the quantum numbers $J^{PC} = 1^{+-}$ (see Eqs. (18-21) in \cite{Nieves:2012fk}). To study this state, a possible current that can be used is:
\begin{eqnarray}
j_{\alpha \beta}^q = \bar Q^a \gamma_\alpha \gamma_5 Q^b \bar q^b \gamma_\beta q^a \label{current}
\end{eqnarray}
where a $\gamma_5$ is inserted into the heavy quark sector so as to change the C-parity to $C=-1$ (it also changes the $P$-parity). The analysis of this $C=-1$ current is similar to the analysis of the $C=+1$ current. In the heavy quark limit, the correlation function using this current can be written as:
\begin{eqnarray}
\tilde \Pi_{\alpha \beta; \gamma\delta } &=& \mbox{Tr} \left[ \gamma_\alpha \gamma_5 \frac{1+\not\!v}{2} \gamma_\gamma \gamma_5 \frac{1-\not\!v}{2}  \right]
\left( {\cal R}_1 g_{\beta \delta} + {\cal R}_2 v_\beta v_\delta \right)
\end{eqnarray}
where the functions ${\cal R}_1$ and ${\cal R}_2$ are identical to the functions appearing in Eq. (\ref{factorgamma}). 
This follows due to the fact that, the currents used for the $C=+1$ and $C=-1$ cases differ only in the structure of the heavy degrees of freedom and are identical in the light degrees of freedom.

Carrying out steps similar to the analysis of the $C=+1$ case, leads to the result that ${\cal R}_1$ has poles when $kv$ is equal to binding energy of the $1^{+-}$ state, and ${\cal R}_1 + {\cal R}_2$ has poles when $kv$ is equal to the binding energy of the $0^{--}$ state. Comparing with the results presented in  Eq. (\ref{aresults2p}), it is seen that $\Lambda_{0^{++}} = \Lambda_{1^{++}}=\Lambda_{2^{++}} = \Lambda_{1^{+-}}$ and also $\Lambda_{0^{--}} = \Lambda_{1^{-+}}$. 

In principal, to study the $C=-1$ states, one can also use the current
\begin{eqnarray}
j_{\alpha \beta} = \bar Q^a \gamma_\alpha  Q^b \bar q^b \gamma_\beta \gamma_5 q^a \label{currentcm}
\end{eqnarray}
i.e. $\gamma_5$ is inserted into the light sector. As the structure of the light degrees of freedom is modified, the correlation function of this current can not be expressed in terms of the functions ${\cal R}_1$ and ${\cal R}_2$. Nevertheless, since the decomposition given in Eq. (\ref{factorgamma}) only uses the structure of the heavy degrees of freedom, a similar decomposition can also be made for this current using different Lorentz invariant functions. In the identification of the spectrum of particles created by this current, the only difference will be that the particles will have the opposite $C$ and $P$ parities. Hence, the results of the previous analysis can be immediately applied to this case:
$\Lambda_{0^{--}} = \Lambda_{1^{--}} = \Lambda_{2^{--}}$. To compare with states of positive $C$ parity, the current
\begin{eqnarray}
j_{\alpha \beta} = \bar Q^a \gamma_\alpha  \gamma_5 Q^b \bar q^b \gamma_\beta \gamma_5 q^a 
\label{current2}
\end{eqnarray}
can be used, leading to the degeneracies $\Lambda_{0^{--}} = \Lambda_{1^{-+}}$ and also $\Lambda_{0^{++}} = \Lambda_{1^{+-}}$ (when comparing results from the currents defined in Eqs. (\ref{currentcm}) and (\ref{current2}) that involve the same structure in the light sector). Note that the last degeneracy corresponds to the $0^{++}$ and $1^{+-}$ states in Eqs. (18-19) of 
\cite{Nieves:2012fk}, which are not degenerate with the $1^{++}$ and $2^{++}$ states.
The distinction between the $0^{++}$ and $1^{+-}$ states that are degenerate with the $1^{++}$ and $2^{++}$ is 
in the light quarks. The states degenerate with $1^{++}$ and $2^{++}$ have their light quarks 
in a state with quantum numbers $J^{PC}=1^{--}$. The others have their light quarks in the state with quantum numbers
$0^{-+}$. 
In the choice of the currents,
$\bar q \gamma_\alpha  \gamma_5 q$ can create a quark-anti-quark pair in the $J^{PC}=0^{-+}$ state whereas $\bar q \gamma_\alpha  q$  creates a pair in the $1^{--}$ state.
For the same reason, the $0^{--}$ and $1^{-+}$ states related to the currents of  Eqs. (\ref{currentcm}) and (\ref{current2}) do not correspond to those discussed for the currents of Eqs. (\ref{current1}) and (\ref{current})

\section{$I=0$ and $I=1$ Components of $X(3872)$}
\label{isospin}
In this section, possible $I=1$ admixture of the $X(3872)$ will be analyzed using the method of \cite{PhysRevD.83.016008}. Below we briefly outline the method.

Let us denote the normalized I=0 and I=1 components of $X(3872)$ state by $\vert X(0)\rangle$ and $\vert X(1) \rangle$. 
The currents that couple to the isospin states $\vert X(0) \rangle$ and $\vert X(1)\rangle$ can be written as
\begin{eqnarray}
j^{I=0} &=& \frac{1}{\sqrt2} \left( j^u + j^d \right)
\nonumber \\
j^{I=1} &=& \frac{1}{\sqrt2} \left( j^u - j^d \right)
\end{eqnarray}
respectively, where $j^q$ is any of the currents given in Eqs. (\ref{spin2current}) and (\ref{spin1current}). 
Let us define the correlation functions
\begin{eqnarray}
\Pi^{qq'} &=& i \int d^4x e^{ipx} \langle 0 \vert {\cal T} j^q(x) j^{q'\dagger}(0) \vert 0\rangle
\nonumber \\
\Pi^{II'} &=& i \int d^4x e^{i p x} \langle 0 \vert {\cal T} j^I(x) j^{I'\dagger}(0) \vert 0 \rangle
\end{eqnarray}
Then
\begin{eqnarray}
\Pi^{00} &=& \frac12 \left( \Pi^{uu} + \Pi^{dd} + \Pi^{ud} + \Pi^{du} \right)
\nonumber \\
\Pi^{11} &=& \frac12 \left( \Pi^{uu} + \Pi^{dd} - \Pi^{ud} - \Pi^{du} \right)
\nonumber \\
\Pi^{10} &=& \frac12 \left( \Pi^{uu} - \Pi^{dd} + \Pi^{ud} - \Pi^{du} \right)
\nonumber \\
\Pi^{01} &=& \frac12 \left( \Pi^{uu} - \Pi^{dd} - \Pi^{ud} + \Pi^{du} \right)
\end{eqnarray}

Since isospin is not an exactly conserved quantity, the states $X(0)$ and $X(1)$ cannot be eigenstates. Hence, they can evolve one into another, i.e. oscillate. This is reflected in the fact that the off diagonal correlation functions are not zero:  $\Pi^{10} \neq 0$ and $\Pi^{01} \neq 0$.

The physical $X(3872)$ and its orthogonal state can be written as:
\begin{eqnarray}
\vert X(3872) \rangle &=& \cos \theta \vert X(0) \rangle + \sin \theta \vert X(1) \rangle
\nonumber \\
\vert X_\perp \rangle &=& - \sin \theta \vert X(0) \rangle + \cos \theta \vert X(1) \rangle  
\label{mixingangledefinition}
\end{eqnarray}
where any possible relative phase can be absorbed in the definition of the states. The respective interpolating currents are given by:
\begin{eqnarray}
j_{X(3872)} &=& \cos \theta j^{I=0} + \sin \theta  j^{I=1} 
\nonumber \\ 
j_{X_\perp} &=& - \sin \theta  j^{I=0} + \cos \theta j^{I=1}
\end{eqnarray}

Being physical eigenstates $\vert X(3872) \rangle$ and $\vert X_\perp \rangle$ should not oscillate, i.e. 
\begin{eqnarray}
&&i \int  d^4 x e^{i p x} \langle 0 \vert {\cal T} j_{X(3872)} (x) j_{X_\perp}^\dagger(0) \vert 0 \rangle 
\nonumber \\ &&
+ i \int d^4xe^{i p x} \langle 0 \vert {\cal T} j_{X_\perp}(x) j_{X(3872)}^\dagger (0) \vert 0 \rangle = 0
\end{eqnarray}
the solution of which gives
\begin{eqnarray}
\tan 2 \theta &=& \frac{\Pi^{10}  + \Pi^{01}}{\Pi^{00} - \Pi^{11}}
= \frac{\Pi^{uu} - \Pi^{dd}}{\Pi^{ud} + \Pi^{du}}
\label{mixingangle}
\end{eqnarray}
In our case,  the correlation functions  that appear in Eq. (\ref{mixingangle}) are either ${\cal R}_1$ or ${\cal R}_2$. It is also implied  that the contributions of states other than $X(3872)$ are subtracted. 
Note that the numerator is non-zero only if isospin is violated, and the denominator receives contributions only from annihilation diagrams. In the annihilation diagrams, the $u(d)$ quarks inserted into the vacuum by the current annihilate into gluons or  photons which then form $d(u)$ quarks that are annihilated by the other current. Such annihilation diagrams are usually omitted in sum rules and lattice calculations since they are considered to be small.
From Eq. (\ref{mixingangle}),
it is seen that if the annihilation diagrams are negligible and/or much smaller than isospin breaking effects, the mixing of the $I=0$ and $I=1$ components in  $X(3872)$ as defined in Eq. (\ref{mixingangledefinition})  can be large (in the limiting case, if one neglects the annihilation diagrams, one obtains maximal mixing as long as there exists isospin symmetry breaking independently of how small the breaking is).

To obtain an order of magnitude estimate of the mixing angle $\theta$, let us first consider the numerator. The numerator receives contributions only from isospin breaking effects. There are two important sources of isospin breaking: the mass differences of the $u$ and $d$ quarks, and electromagnetic interactions. 
From dimensional analysis, one would expect 
$\Pi^{uu} - \Pi^{dd}$ to be of the order of $(m_u - m_d) \Lambda^7$ or $\alpha \Lambda^8$ where $\alpha$ is the electromagnetic coupling and $\Lambda$, in the framework of QCD sum rules, is the Borel mass, which, in the heavy quark limit is of the order of the mass difference between the $X(3872)$ meson and the heavy $c \bar c$ pair, i.e. $\Lambda \sim m_X - 2 m_c = 1.32$ GeV. For such a value of $\Lambda$,  $\alpha \Lambda^8 \sim (m_d - m_u)\Lambda^7$. Both sources give the same order of magnitude contribution to the numerator.  

Considering the denominator in Eq. (\ref{mixingangle}), there is no a priori symmetry limit in which it vanishes. 
Hence, the denominator is ${\cal O}(\Lambda^8)$. Note that, the denominator is also responsible for the difference of the correlation functions $\Pi^{00}$ and $\Pi^{11}$, i.e. it is the term responsible for the mass difference between the $I=0$ and $I=1$ states. If this mass difference is zero, then the denominator should also be zero. Therefore, the denominator can be estimated as $\beta \Lambda^8$ where $\beta$ is a parameter that measures the splitting between the $I=0$ and $I=1$ states. Inserting such a factor $\beta$ is also consistent with the fact that in the hypothetical limit where the splitting between the $I=0$ and $I=1$ states would go to infinity, the mixing angle would go to zero.

Explicit calculation of the parameter $\beta$ for $X(3872)$ is beyond the scope of this letter, but an estimate of it can be obtained from the $\rho$/$\omega$ system. In the currents that are used, the light quarks (which are responsible for the splitting between the $I=0$ and $I=1$ states) are put in a vector configuration, just like in the $\rho$ and $\omega$ mesons.  An estimate of the $\beta$ parameter can be obtained as 
\begin{eqnarray}
\beta \sim \frac{m_\omega - m_\rho}{m_\omega + m_\rho} = 0.0046
\end{eqnarray}

Combining these estimates, the mixing angle can be obtained as:
\begin{eqnarray}
\tan 2\theta = \frac{m_u - m_d}{\beta \Lambda} \simeq 0.56 \longrightarrow \theta \simeq 15^\circ
\label{estimate}
\end{eqnarray}
and hence,
\begin{eqnarray}
\vert X(3872) \rangle \sim 0.96 \vert X(0)\rangle + 0.25 \vert X(1) \rangle
\label{xstates}
\end{eqnarray}

In the models where $X(3872)$ are dominantly a $D D^*$ molecule, the $\vert X(0)\rangle$ and $\vert X(1) \rangle$ components can be written as
\begin{eqnarray}
\vert X(0) \rangle &=& \frac{1}{\sqrt2} \left( \vert D^0 \overline{D^{*0}} \rangle + \vert D^+ D^{*-} \rangle \right)
\nonumber \\
\vert X(1) \rangle &=& \frac{1}{\sqrt2} \left( \vert D^0 \overline{D^{*0}} \rangle - \vert D^+ D^{*-}  \rangle \right)
\label{isospinstates}
\end{eqnarray}
where a $C=+1$ combination is implied. Inserting these into Eq. (\ref{xstates}), the probabilities for the charged and the neutral components can be obtained.
The probability of a molecule of neutral $D$ and $D^*$ meson is $~75\%$, whereas  the probability is $~25\%$ for the charged channel. These results are consistent with the bulk part of the results obtained in \cite{Ortega:2010qq}. (Note that, in 
\cite{Ortega:2010qq}, a possible $c \bar c$ component of $X(3872)$ is also considered).

In \cite{Gamermann:2009uq}  $X(3872)$ is also described as a $D \bar D^*$ molecule. In the notation of 
\cite{Gamermann:2009uq}, the mixing angle $\theta$, defined in Eq. (\ref{mixingangledefinition}), can be written as (see Eq. (127) of \cite{Gamermann:2009uq})
\begin{eqnarray}
\tan^2 \theta = \frac{\int d^3 r \vert \Psi_1(\vec r) - \Psi_2(\vec r) \vert^2}{\int d^3 r \vert \Psi_1(\vec r) + \Psi_2(\vec r) \vert^2} 
\end{eqnarray}
where $\psi_1$($\psi_2$) is the $D^0 \bar D^{*0}$ ($D^+ \bar D^{*-}$) neutral (charged) component of the $X(3872)$ molecular wave-function \cite{Gamermann:2009uq}. From the model of \cite{Gamermann:2009uq}, we obtain $\tan \theta = 0.64$, when a sharp cut-off is used to regularize the ultraviolet divergence. 
This value gives a mixing angle of $\theta \simeq 39^\circ$, which is twice as large as the order of magnitude estimate Eq. (\ref{estimate}) and quite close to maximal mixing $\theta_{max} = 45^\circ$.

Note that the observed isospin violation in the amplitudes of the decays $X(3872) \rightarrow \rho J/\Psi$, and $X(3872) \rightarrow \rho J/\Psi$ is not determined only by the mixing angle, but also by the relative magnitudes of the amplitudes $\langle J/\Psi \rho\vert H \vert X(1)\rangle$ and $\langle J/\Psi\omega \vert H \vert X(0) \rangle$ where we neglect possible isospin violation in decay. The ratio of the amplitudes will be given by\footnote{To evaluate the ratio of the widths, one should
also take into account the phase space of the $\rho$-decaying into two pions and $\omega$ decaying into three pions.}
\begin{eqnarray}
\frac{A(X(3872) \rightarrow J/\Psi \rho)}{A(X(3872) \rightarrow J/\Psi \omega)} 
&=& \tan \theta \frac{\langle J/\Psi\rho \vert H \vert X(1) \rangle}{\langle J/\Psi \omega\vert H \vert X(0)\rangle}
\nonumber \\
&=& \left( \frac{\hat \psi_1 - \hat \psi_2}{\hat \psi_1 + \hat \psi_2} \right)
\end{eqnarray}
where in the last equality, we used the result of \cite{Gamermann:2009uq}, with $\hat \psi_i$  a weighted average of the wave function component $\psi_i(\vec r)$ with a weight that is strongly peaked at origin (zero relative distance between the two mesons). Isospin plays a relevant role in strong processes which are sensitive to short distance dynamics. 
In the molecular picture of $X(3872)$, at short distances, the dynamics of the $X(3872)$ is such that the probability amptitudes of both the neutral and charged meson channels are very similar. This suggests that, when dealing with strong processes, only isospin $I=0$ component will be relevant. The observed isospin breaking in the amplitudes will be small even if the probability to find the $D^0 \bar D^{*0}-$c.c. component in the full space is much larger that that for the $D^+ D^{*-}-$c.c. component.
In \cite{Gamermann:2009uq}, it is shown that this ratio is consistent with the experimental value.

Although the strong decays of $X(3872)$ will mainly be determined by the wave function at the origin, and hence conceal the large isospin violation, the largeness of the mixing angle $\theta$ can lead to significant contributions from the isospin-1 component of $X(3872)$ to processes that are sensitive to large separation between the $D$ mesons making the $X(3872)$ state. Electromagnetic decays such as $X \rightarrow \gamma \Psi(2S)$ which is observed with a branching ratio larger than $3\%$ \cite{pdg12}, will be sensitive to physics at  distances of the order of the size of the $D$-mesons that form the $X(3872)$ state. Possible weak decays of $X(3872)$ in which the $c$-quark decays weakly into an $s$-quark, will be sensitive to even larger separations of the $D$ mesons in the molecular picture. If these weak decays are semi-leptonic, they will also conserve isospin.

\section{Conclusions}
In this letter, even $C$-parity currents given in Eqs. (\ref{spin2current}) and (\ref{spin1current}) that can be used to study $X(3872)$ meson have been proposed. These currents can be used in future for more quantitative analysis of these mesons and a QCD SR analysis is under way. Compared to other currents used in the literature to study the $X(3872)$ mesons (see e.g. \cite{Liu:2008fh,Albuquerque:2012rq}), the proposed currents have the advantage that they can also be used to study the partners of the $X(3872)$ meson on an equal footing.
 
Using the proposed currents, it is proven that the states that couple to them form degenerate triplets with the quantum numbers $J^{PC}=2^{++}$, $J^{PC}=1^{++}$, and $J^{PC}=0^{++}$. Note that since the results are exact in the heavy quark limit, this conclusion holds for {\em any} state that couples to the currents independent of its internal structure. One such triplet is the triplet of mesons $\chi_{c0}$, $\chi_{c1}$ and $\chi_{c2}$. The masses of these particles differ from their average by at most $80$ MeV. Another example is the $\chi_{b0}$, $\chi_{b1}$ and $\chi_{b2}$. In this case, the variation is less then $30$ MeV consistent with an 
$1/m_Q$ effect. Taking these deviation as a measure of the possible $1/m_Q$ effects, $X(3872)$ should have 
spin-0 and spin-2 partners that have a mass that differs by $\sim100$ MeV from the mass of $X(3872)$. We have also analyzed odd $C$-parity currents and end up with a degeneracy spectrum compatible with that derived in the molecular picture of \cite{Hidalgo-Duque:2013fk,Nieves:2012fk}.

Possible existence of a  $I=1$ component in the state $\vert X(3872)\rangle$ is also discussed. It is shown that the mixing angle between the $I=0$ and $I=1$ components  can be large and even close to maximal  \cite{Gamermann:2009uq}. But, it might not be reflected in its strong decays.
Nevertheless, this mixing can be important in decays of $X(3872)$ which are not strong.

\section*{Acknowledgments}
This work is partly supported by the Spanish Ministerio de Economía y Competitividad and European FEDER funds under the contract number FIS2011-28853-C02-01 and FIS2011-28853- C02-02, and the Generalitat Valenciana in the program Prometeo, 2009/090. We acknowledge the support of the European Community-Research Infrastructure Integrating Activity Study of Strongly Interacting Matter (acronym HadronPhysics3, Grant Agreement n. 283286) under the Seventh Framework Programme of the EU. We also acknowledge support from TUBITAK under project number 111T706.

\label{conclusion}
\bibliography{refs2}
\end{document}